\documentclass[aps,prl,twocolumn,showpacs]{revtex4}
\usepackage{amssymb}
\usepackage{amsmath, amsthm}

\begin{document}


\title{Matter instability in modified gravity}


\author{Valerio Faraoni}
\email[]{vfaraoni@ubishops.ca}
\affiliation{Physics Department, Bishop's University\\
Sherbrooke, Qu\`ebec, Canada J1M~0C8
}


\begin{abstract} The Dolgov-Kawasaki instability discovered in 
the matter sector of the modified gravity scenario incorporating 
a $-\mu^4/R$ correction to Einstein gravity is studied in 
general $f(R)$ theories. A stability condition is found in  
the metric version of these theories to help 
ruling out models that are unviable from the theoretical point 
of view.  For example, the theories $f(R)=R+\alpha 
\mu^{2(n+1)}/R^n$, where $\alpha $ and $n$ are real constants 
and $n>0$, are ruled out for any negative value of $\alpha $. 
\end{abstract}


\maketitle

\setcounter{equation}{0}

Analyses of type Ia supernovae  \cite{SN} and of the cosmic 
microwave background, including 
the 3-year WMAP results \cite{WMAP3}, confirm that the present expansion of the universe is 
accelerated. This is 
currently  
explained by invoking a form of dark energy comprising 70\% of the total cosmic energy 
density $\rho$ and with exotic properties (it does not cluster at small scales and has 
negative pressure $P_{de}\simeq -\rho_{de}$). Recently, as an alternative to this exotic dark 
energy, it has been proposed to explain the cosmic acceleration with geometry by modifying 
Einstein gravity at the largest scales (or at low curvatures) by introducing corrections to 
the 
Einstein-Hilbert Lagrangian (``$f(R)$'', or ``fourth order'', or ``modified'' gravity). The 
simplest form for the action of 
modified gravity is 
\begin{equation} \label{1}
S=\left[ \frac{1}{2\kappa}\int d^4x \, \sqrt{-g} \, f(R) 
\right] +S^{(m)} \;,
\end{equation}
where $\kappa \equiv 8\pi G$, $R$ is the Ricci curvature 
\cite{footnoteadded}, and $S^{(m)}$ is the matter action. 
One can vary the action with respect to the metric (``metric 
formalism'') or perform a Palatini variation in which the metric and the connection are 
treated as independent variables \cite{Palatini} (``Palatini formalism'')---the resulting 
field equations, 
which coincide in general relativity, are different in higher 
order gravity. Furthermore, if the matter  action $S^{(m)}$ also 
depends on the connection, one 
obtains a third possibility, metric-affine gravity theories 
\cite{LiberatiSotiriou}.

 Quadratic quantum corrections were originally introduced to improve the 
renormalizability of general relativity  \cite{renorma} and were 
used to achieve inflation in 
the early universe \cite{Starobinsky}.  There is also 
motivation for modified gravity from 
string/M theory \cite{stringmotivations}.

The field equations obtained by varying the action~(\ref{1}) in the metric formalism are 
\begin{equation} \label{1bis}
f' R_{ab}-\frac{f}{2}\, g_{ab}=\nabla_a\nabla_b f'-g_{ab} \, \Box 
f'+\kappa \, T_{ab} \;,
\end{equation}
where  a prime denotes differentiation with respect to $R$, $g_{ab}$ and $R_{ab}$ are the 
metric and Ricci tensors, and $T_{ab}$ is the stress-energy 
tensor of ordinary matter.  The idea of modified gravity consists in introducing 
corrections 
to the Einstein-Hilbert action that are negligible in the early universe and only become 
effective 
at late times when the Ricci curvature $R$ decreases (the radiation era with $R=0$ deserves 
a special discussion --- see below). The prototype is the theory $f(R)=R-\mu^4/R$ 
\cite{MG1,CDTT,oneoverR} which, however, was found to be 
unstable 
\cite{DolgovKawasaki,Odintsovconfirm}. Therefore, we parametrize the deviations from Einstein 
gravity  as
\begin{equation}\label{1ter}
f(R)=R+ \epsilon \, \varphi(R) \;,
\end{equation}
where $\epsilon $ is a small parameter with the dimensions of an inverse squared length and 
$\varphi$ is arranged to be dimensionless (in the previous example with $f=R-\mu^4/R$ it is 
$\epsilon=-\mu^4$ with $\mu\simeq H_0\approx 10^{-33}$~eV, the present value of the Hubble 
parameter $H$). In order for the effects of these 
corrections to appear only at recent times a sensible choice is $\epsilon \approx H_0^2\simeq 
10^{-66}\; \mbox{eV}^2$, which amounts to some fine-tuning which is present in all these 
theories, as well as in dark energy models. At 
present there is little indication on what the correct modification of Einstein gravity 
should be  and it would be useful to weed out some of the competing models. Besides 
compatibility with experimental data (e.g., \cite{Menaetal}), 
minimal criteria 
that a 
modified gravity theory must satisfy in order to be viable are the following: 1) 
reproducing the desired dynamics of the universe including an inflationary era, followed by a 
radiation era (required and well-constrained by primordial 
baryogenesis and nucleosynthesis \cite{LambiaseScarpetta} 
and a matter era (required for the formation of structures) and, finally, by the present 
acceleration epoch (constrained by cosmic microwave background experiments, supernova data,  
and large scale 
structure surveys). There is recent evidence 
\cite{AmendolaPolarskiTsujikawa} (see also 
\cite{Amendolacomments}) that 
the popular models $f(R)=R-\mu^{2(n+1)}/R^n$ with $\mu, n>0$ 
\cite{MG1,CDTT,oneoverR} do not allow a 
matter-dominated era and are therefore ruled out. 2) Having Newtonian and 
post-Newtonian limits compatible with the available Solar System experiments. This 
issue is not settled in general, there are conflicting results on the weak-field limit 
of modified gravity in specific scenarios, and we still lack a 
general formalism applicable 
to all  modified gravity models \cite{PPN,IgnacioKarel,veto}. 3) 
The theory must 
be stable at the  classical and quantum level. There are in principle several kinds of 
instabilities to consider \cite{instabilities}. Dolgov and 
Kawasaki  \cite{DolgovKawasaki} 
discovered an 
instability in the matter sector for the specific model 
$f(R)=R-\mu^4/R$, a result confirmed 
in  Ref.~\cite{Odintsovconfirm}, in which it is also shown that 
adding a term quadratic in $R$ removes the instability. 
Instabilities of de Sitter space in the gravity sector were 
found in Refs.~\cite{mypapers,DolgovPelliccia, 
BarrowHervik,MullerSchmidtStarobinsky}, 
while stability with respect to black hole nucleation was 
studied in Refs.~\cite{Cognolaetal}. The consideration of 
physically 
different instabilities yields remarkably similar stability conditions. The theory should 
also be 
ghost-free: the presence of ghosts 
has been studied in Refs.~\cite{IgnacioKarel}. In this 
communication we focus on the  Dolgow-Kawasaki instability and 
generalize  their analysis to arbitrary 
$f(R)$ theories.

By taking the trace of eq.~(\ref{1bis}) one obtains
\begin{equation} \label{2}
3\Box f'+ f' R -2f=\kappa T \;.
\end{equation}
By evaluating $\Box f'$ this equation is written as
\begin{equation}\label{3}
\Box R +\frac{\varphi '''}{\varphi ''}\nabla^a R \, \nabla_a R+\frac{\left( \epsilon \varphi 
'-1\right) }{3\epsilon \, \varphi ''}\, R=\frac{\kappa \, T }{3\epsilon 
\, \varphi''}\, +\, \frac{\varphi}{3\varphi ''} \;.
\end{equation}
We assume that the second derivative $\varphi '' $ is different 
from zero; if $\varphi ''=0$ on an interval then the theory 
reduces to general relativity and is not of interest here. 
Isolated zeros of $\varphi ''$ (at which the theory is 
``instantaneously general relativity'') are  not dealt 
with in this paper, but they are generally allowed and the 
following discussion applies to the distinct regions 
in which $\varphi''$ maintains its sign.

We now follow the procedure of Ref.~\cite{DolgovKawasaki} and 
consider a small region of 
spacetime in the weak-field regime, in which the metric and the curvature can locally be 
approximated by
\begin{equation}\label{4}
g_{ab}=\eta_{ab}+h_{ab} \;, \;\;\;\;\;\;\;\;\;\;\;\;\; R=-\kappa\, T +R_1 \;,
\end{equation}
where $\eta_{ab} $ is the Minkowski metric and $\left| R_1/\kappa\, T\right| \ll 1$. This 
inequality excludes the case of conformally 
invariant matter with $T=0$, including the important case of  a radiation fluid 
with equation of state $P=\rho/3$ --- this situation is examined later. 
Equation~(\ref{4}) yields, to first order in $R_1$,
\begin{eqnarray}
&& \ddot{R}_1 -\nabla^2 R_1 -\frac{2\kappa\, \varphi '''}{\varphi ''}\, \dot{T}\dot{R}_1+\, 
\frac{2\kappa \, \varphi '''}{\varphi ''}\, \vec{\nabla}T \cdot \vec{\nabla}R_1 \nonumber \\
&& \nonumber \\
&& +\frac{1}{3\varphi ''} \left( \frac{1}{\epsilon}-\varphi' \right) R_1=
\kappa \, \ddot{T}-\kappa \nabla^2 T -\, \frac{\left(\kappa\, T\varphi '+\varphi 
\right)}{3\varphi ''}\;,\nonumber \\
&& \label{5}
\end{eqnarray}
where $\vec{\nabla}$ and $\nabla^2$ denote the gradient and Laplacian operators in Euclidean  
three-dimensional space, respectively, and an overdot denotes differentiation with respect 
to time. The function $\varphi$ and its derivatives are now evaluated at $R=-\kappa\, T$. The 
coefficient of $R_1$ in the fifth 
term on the 
left hand side is the square of an effective mass and is dominated by the term  $\left( 
3\epsilon \, 
\varphi '' \right)^{-1}$ due to the 
extremely small value of  $\epsilon$ needed for these theories to reproduce the correct 
cosmological dynamics. It is therefore obvious that 
the theory will be stable if $\varphi '' =f''>0$, while an instability arises if this 
effective mass is negative, i.e., if $\varphi''=f''<0$. The time scale for this instability 
to manifest is estimated to be of order $10^{-26}$~s for the specific case 
$\varphi(R)=-\mu^4/R$ \cite{DolgovKawasaki}. As shown by this example,  the 
smallness of $\varphi ''$ contributes to the magnitude of the effective mass and the 
smallness of the characteristic time scale for the instability to develop. Although a 
$\varphi ''$ term also appears in the denominator of the last source term on the right hand 
side of 
eq.~(\ref{5}), it is partially offset by the presence of $\varphi$ and $\varphi'$ in the 
numerator, while this does not occurs for the dominant term in the effective mass of $R_1$.

Let us consider as an example the  model  
$f(R)=R-\frac{\mu^{2(n+1)}}{R^n}$ with $\mu>0$ 
and $n>0$ ($n$ is not restricted to be an integer): since 
$f''=-n\left(n+1 
\right)\mu^{2(n+1)}/R^{n+2} <0$ this theory always suffers 
from the Dolgov-Kawasaki instability. Instead, the theory
$f(R)=R+\frac{\mu^{2(n+1)}}{R^n}$ (again with $\mu, n>0$) is stable (but the 
required cosmological dynamics may not be achieved).
As another example consider $f(R)=R+\Lambda +\alpha R^n$ with $n>0$ and 
$\Lambda>0$ a cosmological constant, corresponding to 
Starobinsky inflation in the early universe \cite{Starobinsky}. The $R^2$ correction 
to the Einstein-Hilbert Lagrangian is 
large during this early epoch and the contribution of matter to the dynamics is negligible; 
any instability  could in principle be interesting, and  
$f''=n\alpha R^{n-2} >0$ if $\alpha>0$, corresponding to stability. 

We can now consider the case of  a radiation fluid or other form of matter with 
vanishing trace $T$ of the stress-energy tensor. In this case eq.~(\ref{5}) becomes
\begin{eqnarray}
&&\ddot{R}_1+\frac{\varphi '''}{\varphi ''} \, \dot{R_1}^2 -\nabla^2 R_1 
-\frac{\varphi '''}{ \varphi ''} \, \left(\vec{\nabla} R_1 \right)^2 \nonumber \\
&&\nonumber \\
&& +\frac{1}{3\varphi 
''}\left( \frac{1}{\epsilon}-\varphi '\right) R_1 
 =\frac{\varphi}{3\varphi ''} \;.\label{6}
\end{eqnarray}
Again, the effective mass term is $\approx \left( 3\epsilon 
\varphi '' \right)^{-1}$, which  has the sign of $f''$ and the 
previous stability criterion is recovered \cite{footnoten}.

One could think of using an independent approach and  study the equilibrium 
non-perturbatively by taking advantage of the dynamical equivalence between $f(R)$ gravity 
and a scalar-tensor theory. By introducing the auxiliary field $\phi$, the action (\ref{1}) 
can be rewritten as
\begin{equation} \label{13}
S=\frac{1}{2\kappa} \int d^4 x \, \sqrt{-g}\,\left[ \psi(\phi) R-V(\phi) \right] +S^{(m)} 
\end{equation}
when $f''\neq 0$, where
\begin{equation} \label{14}
\psi(\phi)=f'(\phi) \;, \;\;\;\;\;\;\;\;\;\; V(\phi)=\phi f'(\phi) -f(\phi) \;.
\end{equation}
The corresponding field equations are
\begin{equation} \label{15}
G_{ab}=\frac{1}{\psi} \left( \nabla_a\nabla_b \psi -g_{ab} \Box \psi -\frac{V}{2} g_{ab} 
\right) +\frac{\kappa}{\psi}\, T_{ab} \;,
\end{equation}

\begin{equation}\label{16}
R\, \frac{d\psi}{d\phi}-\, \frac{dV}{d\phi}=0 \;.
\end{equation}
The action (\ref{13}) reduces to  (\ref{1})  trivially if $\phi=R$ and, vice-versa, 
variation 
of (\ref{13}) with respect to $\phi$ yields $\phi=R$ if $f''\neq 0$. Now one may think 
of studying the 
stability 
of the model by looking at the shape and extrema of the effective potential $V(\phi)$ but 
this would be misleading. In fact, the dynamics of $\phi$ are not regulated by $V$ but are 
fixed by the strong constraint $\phi=R$. By taking the trace of eq.~(\ref{15}) and using 
eq.~(\ref{14}), one obtains again eq.~(\ref{2}). Equation~(\ref{16}) for $\phi$ has no 
dynamical 
content because, using eq.~(\ref{14}) with $\phi=R$, it is identically satisfied. This is 
linked to the fact that the kinetic term for $\phi$ is absent in the action (\ref{13}) with 
Brans-Dicke parameter $\omega=0$ and the dynamical equation for $\phi$ degenerates into an 
identity. Similar considerations apply to the Palatini action, which is 
equivalent to a 
Brans-Dicke theory  with parameter $\omega=-3/2$.

In conjunction with the other theoretical requirements listed above, the (in)stability 
condition derived here constitutes a useful criterion to  veto 
some of the proposed $f(R)$ 
gravity scenarios  from a theoretician's point of view, in order  to focus on more promising 
models.

Note that, when $f''<0$, the instability of these theories can be interpreted, following 
eq.~(\ref{5}), as an instability in the gravity sector. Equivalently, by using eq.~(\ref{4}), 
it can be seen as a matter instability (this is the interpretation taken in 
\cite{DolgovKawasaki}). Whether the instability arises in the gravity or matter sector seems 
to be  a matter of interpretation.

A problem potentially related to the instability is the 
weak-field limit of $f(R)$ gravity,  which has been the 
subject of several papers reporting conflicting (sometimes even 
opposite) results \cite{PPN,IgnacioKarel,veto}. We hope 
to  report soon on  this issue and its relation with local 
instability.

Finally, we mention the issue of the Cauchy problem: a  well 
posed Cauchy problem is necessary if the theory is to have 
predictive power  \cite{Wald}. It was shown in 
Ref.~\cite{Noakes}, using harmonic coordinates, that theories 
of the form
\begin{equation}
S=\int d^4x \, \sqrt{-g}\left( R+\alpha R_{ab}R^{ab}+\beta R^2 
\right)
\end{equation}
have  a well posed initial value problem. Moreover, the 
existence of a well posed Cauchy problem can be reduced to the 
analogous problem by using the dynamical equivalence with 
scalar-tensor gravity (\ref{13}) and (\ref{14}) when $f''\neq 
0$. The well posedness of the Cauchy problem was demonstrated for 
particular scalar-tensor theories in 
Refs.~\cite{CockeCohen,Noakes} and has recently been the subject 
of a thorough analysis \cite{Salgado}. Although further study is 
needed for the specific cases $\omega=0 $ and $\omega=-3/2$, 
which are relevant for $f(R)$ gravity in the metric and Palatini 
formalism, respectively, it seems that the Cauchy problem is 
well posed \cite{Salgado}. Details on this subject will be 
presented elsewhere.

\begin{acknowledgments}
The author thanks the organizers of the 2006 Modern Cosmology 
Workshop in Benasque and the Benasque Center for Science, where 
this work was carried out, T.P Sotiriou for pointing out an 
error in a previous version of this paper, and financial support 
from the 
Natural Sciences and Engineering Research Council of Canada ({\em NSERC}).
\end{acknowledgments}


\begin{thebibliography}{99}
\bibitem{footnoteadded} In our notations, which follow 
Ref.~\cite{Wald}, $R=6\left(  \frac{\ddot{a}}{a}+\frac{ 
\dot{a}^2}{a^2} \right) $ in a 
spatially flat Friedmann-Robertson-Walker universe with scale 
factor $a(t)$, an overdot denoting differentiation with respect 
to the comoving time $t$.
\bibitem{footnoten} A stability condition similar to $f''>0$ was 
obtained in Ref.~\cite{SongHuSawicky} in the metric formalism, 
and for cosmological perturbations in the Newtonian gauge.
\end{thebibliography}

\end{document}